\documentclass[twocolumn,showpacs,preprintnumbers,amsmath,amssymb,prl,superscriptaddress]{revtex4}

\usepackage{amsmath,amssymb,amsthm,color}
\usepackage{graphicx}
\usepackage{dcolumn}

\newcommand {\beq}{\begin{eqnarray}}
\newcommand {\eeq}{\end{eqnarray}}

\begin{document}

\preprint{CALT-TH 2015-037, IPMU 15-0105}

\title{Bulk Locality and Boundary Creating Operators}

\author{Yu Nakayama}

\affiliation{Walter Burke Institute for Theoretical Physics \\ California Institute of Technology, Pasadena, California 91125, USA}

\author{Hirosi Ooguri}

\affiliation{Walter Burke Institute for Theoretical Physics \\ California Institute of Technology, Pasadena, California 91125, USA}
\affiliation{Kavli Institute for the Physics and Mathematics of the Universe \\ 
University of Tokyo, 
5-1-5 Kashiwanoha, Kashiwa, Chiba 277-8583, Japan}


\begin{abstract}
We formulate a minimum requirement for CFT operators to be localized in the dual AdS.  
In any spacetime dimensions, we show that a general solution to the requirement is a linear superposition 
of operators creating spherical boundaries in CFT, with the dilatation by the imaginary unit from their centers. 
This generalizes the recent proposal by Miyaji  {\it et al.} for bulk local operators in the three dimensional AdS. 
We show that Ishibashi states for the global conformal symmetry in any dimensions and with the imaginary dilatation 
obey free field equations in AdS and that incorporating bulk
interactions require their superpositions. We also comment on the recent proposals by
Kabat {\it et al.,} and by H.~Verlinde.

\end{abstract}

\maketitle


Recently, Miyaji {\it et al.} \cite{Miyaji:2015fia} 
proposed a construction of bulk local states and corresponding operators in the three-dimensional 
AdS using Ishibashi states  \cite{Ishibashi:1988kg}, 
which create spherical boundaries in the dual CFT in two dimensions.
To be precise, the states they proposed preserve one half of the global conformal symmetry 
and not of the full Virasoro symmetry. 
A similar but different construction was also proposed by H.~Verlinde \cite{Verlinde:2015qfa}.
In this paper, we will formulate a minimum requirement on bulk local operators in AdS 
and show that its solution generalizes the construction 
by Miyaji {\it et al.}, both in spacetime dimensions and in $1/N$ corrections. We will also comment on 
its relation to the earlier construction using a bulk-boundary kernel in AdS 
\cite{Hamilton:2006az,Kabat:2011rz,Heemskerk:2012mn,Kabat:2013wga,Kabat:2015swa}.

Consider CFT in $d$ dimensions and its dual AdS gravity in $(d+1)$ dimensions. 
A bulk local operator $\hat{\psi}$ is an {\it operator} in the Hilbert space of CFT
and is a {\it local} probe of the {\it bulk} AdS geometry. As a local probe, it must depend 
on the position $(t, \rho, \vec{x})$ 
in AdS with the metric, 
\begin{align}
ds^2 = -\cosh^2\rho \ dt^2 + d\rho^2 + \sinh^2\rho \ d\vec{x}^2 ,
\end{align}
where we set the AdS radius to be $1$ and use $\vec{x}$ subject to $\vec{x}^2 = 1$ in $d$ dimensions  
to parametrize the sphere S$^{d-1}$. 

The only requirement we impose on $\hat{\psi}$ is that the actions of
the conformal symmetry and the bulk isometry on $\hat\psi$ are compatible. 
Namely, 
 \begin{align}
[J, \hat{\psi}(t, \rho , \vec{x}) ] = i\mathcal{L}_{{\mathcal J}} \hat{\psi}(t, \rho , \vec{x}) , \label{inf} 
\end{align}
where $J$ on the left-hand side is a generator of the conformal symmetry of CFT and
 $\mathcal{L}_{{\mathcal J}}$ on the right-hand side is the Lie derivative in the AdS
coordinates $(t, \rho, \vec{x})$
with respect to the Killing vector field ${\mathcal J}$ corresponding to $J$.
We will show that a general solution to (\ref{inf})  is a linear superposition of boundary creating operators 
in CFT with the dilatation by the imaginary unit.
When $\hat{\psi}$ is a scalar field in AdS, the Lie derivative on the right-hand side of (\ref{inf}) acts as,
\begin{align}
\mathcal{L}_{{\mathcal J}} \hat{\psi} = \mathcal{J}^\mu \partial_\mu \hat{\psi}.
\label{scalar}
\end{align}
In general, $\hat\psi$ may carry a spin, {\it i.e.}, belong to
a finite dimensional representation of the isotropy group $SO(1,d)
\subset SO(2,d)$ preserving a point in AdS.

The $SO(2,d)$ conformal generators on the cylinder $\mathbb{R} \times {\rm S}^{d-1}$ can 
be organized as $J=(H,M_{ab}, P_a, K_a)$, where  $a, b=1,\cdots d$, $H$ is the global Hamiltonian along $\mathbb{R}$, 
$M_{ab} = -M_{ba}$ generate rotations on S$^{d-1}$, and $P_a$ and $K_a$ are the translation and the special conformal generators
when  $\mathbb{R} \times {\rm S}^{d-1}$ is mapped onto $\mathbb{R}^d$. 
Their commutation relations relevant in the following
discussion are
\begin{align}
& [K_a, P_b] = 2\delta_{ab} H -2iM_{ab} , \cr
& [H, P_a] = P_a, \ [H, K_a] = -K_a  .
\end{align}
The Belavin-Polyakov-Zamolodchikov conjugation rule is $K_a^\dagger = P_a$. $H^\dagger = H$, $M_{ab}^\dagger = M_{ab}$.

Let us examine implications of the requirement (\ref{inf}) when $\hat \psi$ is located at the origin $\rho=0$ on the
$t=0$ slice. The isotropy group $SO(1,d)$ of this point is generated by 
$M_{ab}$ and $P_a + K_a$. Therefore, if $\hat\psi$ is a scalar field, (\ref{inf}) combined with
(\ref{scalar}) gives,  
\begin{align}
[M_{ab},   \hat{\psi}(0)] &= 0  ,   \cr
[P_a + K_a, \hat{\psi}(0) ] &=  0  .
\label{commute}
\end{align}
More generally, if $\hat\psi$ is in a finite dimensional representation of the isotropy group $SO(1,d)$,
\begin{align}
[M_{ab},   \hat{\psi}(0)] &= s_{ab} \hat{\psi}(0) ,   \cr
[P_a + K_a, \hat{\psi}(0) ] &=   s_{0a}  \hat{\psi}(0) ,
\end{align}
for some matrices $s_{ab} = -s_{ba}$ and $s_{0a}$ characterizing the spin of $\hat \psi$. 

Acting $\hat{\psi}(0)$ on the conformally invariant vacuum $|0 \rangle$, 
we obtain the state $|\psi(0) \rangle \equiv \hat{\psi}(0)|0 \rangle$, which satisfies
\begin{align}
M_{ab}|\psi(0) \rangle & = s_{ab} |\psi(0) \rangle,   \nonumber \\
(P_a + K_a)|\psi(0) \rangle & =  s_{0a} |\psi(0) \rangle.
 \label{orig}
\end{align}
The Hilbert space of CFT is decomposed into a sum 
of irreducible highest weight representations of the conformal algebra.
These equations have a unique solution within each of the 
representations. 

For example, when $\hat{\psi}$ is a scalar field,
the corresponding state $|\psi(0)\rangle$ satisfies, 
\begin{align}
M_{ab}|\psi(0) \rangle & = 0 ,   \nonumber \\
(P_a + K_a)|\psi(0) \rangle & =  0 .
 \label{org5}
\end{align}
We can solve these equations by starting with any conformal primary state $|\phi \rangle$ satisfying,
\begin{align}
  H|\phi\rangle =\Delta_\phi |\phi\rangle \ , ~ 
K_a |\phi\rangle=0 \ , ~  M_{ab} |\phi\rangle = 0 , 
\end{align}
 and by adding its descendants as,
\begin{align}
|\phi \rangle\rangle = \sum_{n=0}^\infty (-1)^n C_n (P^2)^n | \phi \rangle ,
\end{align}
where  $P^2 = \sum_{a=1}^d P_a P_a$ and 
\begin{align}
C_n = \prod_{k=1}^n \frac{1}{4k \Delta_{\phi} + 4 k^2 -2kd} ,
\label{whatisC}
\end{align}
up to an overall normalization independent of $n$. It turns out that
the sum over $n$ can be expressed in term of the Bessel function $J_\nu(x)$ of the first kind as,
\begin{align}
|\phi \rangle\rangle = \Gamma\left(\Delta_{\phi}-\frac{d}{2}+1\right) \left(\frac{\sqrt{P^2}}{2}\right)^{\frac{d}{2}-\Delta_{\phi}} J_{\Delta_{\phi}-\frac{d}{2}} (\sqrt{P^2})|\phi \rangle ,
\end{align}
We note that it is related to the Fourier transform of a bulk-boundary Green function in AdS.
A general solution to \eqref{orig} is then a linear combination of $|\phi \rangle\rangle$ over 
primary states $\phi$,
\begin{align}
  |\psi(0) \rangle = \sum_\phi \psi_\phi |\phi \rangle\rangle . 
\label{superposition}
\end{align}

These states are naturally related to boundary conditions in CFT. 
To see this, perform time evolution on $|\phi \rangle\rangle$ to define a new state, 
$|\phi_{{\rm Ishibashi}} \rangle\rangle = e^{ i \frac{\pi}{2} H} | \phi \rangle\rangle$.
It satisfies,
\begin{align}
M_{ab}|\phi_{{\rm Ishibashi}} \rangle\rangle &=0 , \nonumber \\
(P_a-K_a)|\phi_{{\rm Ishibashi}} \rangle\rangle &=0 , 
\label{conformalboundary}
\end{align}
and is expressed as,
\begin{align}
 |\phi_{{\rm Ishibashi}} \rangle\rangle = e^{i \frac{\pi}{2}\Delta_\phi}
\sum_{n=0}^\infty C_n (P^2)^n | \phi \rangle ,
\label{Ishibashi}
\end{align}
with the coefficients $C_n$ given by (\ref{whatisC}).
We recognize that (\ref{conformalboundary}) are exactly the conditions on 
conformal boundary states located at the equator of the Euclidean S$^d$ related to the 
the $t=0$ slice of the Lorentzian  $\mathbb{R} \times {\rm S}^{d-1}$ by the Wick rotation $t = -i\tau$. 
They preserve $SO(1, d) \subset SO(1,d+1)$ of the Euclidean conformal group.

Our boundary states $|\phi_{{\rm Ishibashi}} \rangle\rangle$,
defined in any dimensions, generalize Ishibashi states for two dimensional CFT  \cite{Ishibashi:1988kg}.
For this reason, we call  $|\phi \rangle\rangle = e^{-i\frac{\pi}{2} H} |\phi_{\rm{Ishibashi}}\rangle\rangle$
as a ``twisted Ishibashi state.'' The global Hamiltonian
$H$ acting on a boundary state is the dilatation operator 
from the center of the spherical boundary 
generated by the state. Therefore, we may interpret 
$ e^{-i\frac{\pi}{2} H}$ as the dilatation by the imaginary unit,
$e^{i \pi/2}=i$. 

When $d=2$, the state $|\phi \rangle\rangle = e^{-i\frac{\pi}{2} H} |\phi_{\rm{Ishibashi}}\rangle\rangle$
reduces to the one proposed by \cite{Miyaji:2015fia} as a bulk state localized at the origin $\rho=0$ on the $t=0$ slice.
Note, however, that a general solution to (\ref{org5}) is a superposition of twisted Ishibashi states
(\ref{superposition}). As we shall see below, bulk interactions and the microscopic causality require a non-trivial superposition since
each twisted Ishibashi state obeys a free field equation in AdS. 

More generally, when the bulk local operator $\hat{\psi}$ carries a non-trivial spin, 
$|\psi(t=-\pi/2) \rangle = e^{i \frac{\pi}{2} H} |\psi(0)\rangle$ 
satisfies, 
\begin{align}
M_{ab}|\psi(t=-\pi/2) \rangle& =s_{ab}|\psi(t=-\pi/2)\rangle , \nonumber \\  
(P_a - K_a) |\psi(t=-\pi/2) \rangle & = is_{0a}|\psi(t=-\pi/2) \rangle . \label{orig3}
\end{align}
A solution to these equations can also be found by starting with a primary state $|\phi \rangle$ in the
same representation $\{ s_{ab}, s_{0a} \}$ of $SO(1,d)$ and by adding conformal descendants, whose coefficients
are fixed by (\ref{orig3}) iteratively, except when there are null vectors in the representation, in which case
the solution is not unique (this correspond to gauge degrees of freedom in the bulk). 
Existence of these states suggests that conformal boundary conditions (\ref{conformalboundary}) can be generalized so that
boundary states are not invariant under the $SO(1,d)$ subgroup of the conformal symmetry but are in its finite dimensional 
representation. It would be interesting to explore their interpretation from the point of view of CFT. 

The relation between the bulk local state $|\psi(0) \rangle$ at the origin of AdS and a boundary
state of CFT can be explained intuitively as follows. 
The conformal generators $M_{ab}$ and $P_a - K_a$, which annihilate boundary states,
generate isometry on the $t=0$ slice of AdS. 
Therefore, if a boundary state is dual to a gas of free massive particles in AdS,  these particles 
must be uniformly distributed at $t=0$.
Now, geodesics in AdS are $2\pi$ periodic in the global time $t$.
If  a  massive particle has zero orbital angular momentum, it comes to the origin at $\rho = 0$
at every half period $\pi$.  
Therefore, the uniformly distributed gas of free massive particles at $t=0$ will 
converge at the origin $\rho=0$ on the $t=\pi/2$ slice, showing
how time evolution of the boundary state by a quarter of the period gives a localized state. 

Once a state $|\psi (0) \rangle$ localized at the origin of AdS is constructed, it can be moved
to an arbitrary point by the AdS isometry. More explicitly,  
we can map the origin $\rho = 0$ to any point $(\rho, \vec{x})$ on the $t=0$ slice 
using the generators $P_a - K_a$, and we can then use $H$ to move to a different time slice as,
\begin{align}
|\psi (t, \rho,\vec{x}) \rangle = e^{-i H t} e^{\rho (P_a - K_a) x^a} |\psi(0) \rangle . \label{result}
\end{align}

Since the AdS coordinates $(t, \rho,\vec{x})$ are coupled to the generators $H$ and $P_a - K_a$ of 
the coset $SO(2,d)/SO(1,d)$
by the exponential map in (\ref{result}), their infinitesimal variations automatically give,
\begin{align}
J |  \psi (t ,\rho,\vec{x})\rangle   = i \mathcal{L}_{\mathcal{J}} | \psi (t ,\rho,\vec{x})\rangle  \label{Ward} , 
\end{align}
for each generator $J=(H,M_{ab}, P_a, K_a)$ of $SO(2,d)$. Therefore, the corresponding operator  $\hat{\psi}(r, \rho, \vec{x})$
satisfies our requirement (\ref{inf}) for bulk local operators.

So far, we have studied solutions to (\ref{org5}) and found that they are linear superpositions of twisted Ishibashi
states as in (\ref{superposition}). We would like to discuss how the superposition 
coefficients $\psi_\phi$ are determined. 

For each primary state $|\phi \rangle$,
\begin{align}
|\phi (t, \rho,\vec{x}) \rangle\rangle = e^{-i H t} e^{\rho (P_a - K_a) x^a} |\phi \rangle\rangle, 
\end{align}
obeys a free field equation in AdS. To see this, note that 
$|\phi (t, \rho,\vec{x}) \rangle\rangle$ is constructed in a single irreducible representation
and therefore is an eigenstate of the quadratic Casimiar operator
of $SO(2,d)$ with the eigenvalue $m^2 = \Delta_\phi (\Delta_\phi-d)$. The compatibility condition (\ref{Ward}) of the conformal symmetry of CFT and
the isometry of AdS then turns the  operator into the Laplace-Beltrami operator on $|\phi (t, \rho, \vec{x})\rangle\rangle$, and the free field
equation with mass $m$ follows.

For example, we can compute an overlap of a scalar primary state $|\phi \rangle$ and the corresponding twisted Ishibashi state $|\phi (t, \rho, \vec{x})\rangle\rangle$
as, 
\begin{align}
\langle \phi |\phi (t, \rho, \vec{x})\rangle\rangle &= \langle \phi| e^{-i H t} e^{\rho (P_a - K_a) x^a}  |\phi \rangle\rangle
\nonumber \\ &=\frac{e^{-i \Delta_\phi t} }{{\cosh\rho}^{\Delta_\phi}}, \label{wave}
\end{align}
which indeed satisfies the Klein-Gordon equation in AdS.

Thanks to the state-operator correspondence of CFT, we can construct an operator $\hat{\phi}(t, \rho, \vec{x})$ for the twisted Ishibashi state
$| \phi(t, \rho, \vec{x})\rangle\rangle$. Could it be a bulk local operator? We can answer this question by computing 
 the two-point function,
 \begin{align}
  \langle 0 | \hat{\phi} (t, \rho, \vec{x}) \hat{\phi} (t', \rho', \vec{x}') |0\rangle = \langle\langle \phi(t, \rho, \vec{x}) |
\phi(t', \rho', \vec{x}')\rangle\rangle.
\end{align}
 Since  $| \phi(t, \rho, \vec{x})\rangle\rangle$ belongs to a highest weight representation, it is a sum of positive energy states. 
Together with the free field equation on $| \phi(t, \rho, \vec{x})\rangle\rangle$ and 
the boundary condition on the two-point function at $t=t'$, the two-point function is uniquely determined to be  
the Wightman function of a free field in AdS (see also appendix B of \cite{Miyaji:2015fia}). If $\hat{\phi}(t, \rho, \vec{x})$ is a truly local operator in the bulk, the Reeh-Schlieder theorem would imply that its higher point correlation functions are trivial \cite{Morrison:2014jha}. Namely, 
$\hat{\phi}(t, \rho, \vec{x})$ would be a free field in the bulk.

At this point, it is instructive to compare our requirement on bulk local operators with the proposal by  Kabat, Lifshytz, and Lowe (KLL) \cite{Hamilton:2006az, Kabat:2011rz, Kabat:2013wga} (see also \cite{Banks:1998dd, Balasubramanian:1998sn, Bena:1999jv} for earlier papers).
Their construction at the leading order in $1/N$ expansion is 
\begin{align}
\hat{\phi}^{\mathrm{KLL}}_0(t, \rho,\vec{x}) = \int dt' dx' K(t, \rho,\vec{x}| t', \vec{x}') \phi(t', \hat{x}') , \label{leading}
\end{align}
where  $\phi(t', \hat{x}') $ is the primary field in CFT corresponding to $|\phi \rangle$. 
The bulk-boundary kernel $K(t, \rho,\vec{x}| t', \vec{x}')$, which is called as a smearing function in \cite{Hamilton:2006az, Kabat:2011rz, Kabat:2013wga}, 
can be extracted from the boundary behavior $\rho' \rightarrow 0$ of a bulk Green's function $G(t, \rho, \vec{x} |t', \rho', \vec{x}')$ as,
\begin{align}
&G(t, \rho, \vec{x} |t', \rho', \vec{x}') \nonumber \\
& \sim \frac{ {\rho'}^{\Delta_\phi} L(t, \rho, \vec{x} |t', \vec{x}') + {\rho'}^{d-\Delta_\phi}K(t,\rho, \vec{x} |t', \vec{x}')}{2\Delta_\phi - d}  .
\end{align} 
We can choose the Green's function $G$ so that
 $K(t, \rho,\vec{x}| t', \vec{x}')$ is non-zero only when  $(t, \rho,\vec{x})$ and  $(t', \vec{x}')$ are
 space-like separated \cite{Hamilton:2006az,Heemskerk:2012mn}.  

We claim that the operator $\hat{\phi}(t, \rho, \vec{x})$ corresponding to the twisted Ishibashi state $|\phi(t, \rho, \vec{x})\rangle\rangle$
is identical to  $\hat{\phi}^{\mathrm{KLL}}_0(t, \rho,\vec{x})$ given by (\ref{leading}).
This follows from the facts that both satisfy the compatibility condition (\ref{inf}) and that both generate states
in the same irreducible representation with the highest weight state $|\phi \rangle$.
Since these conditions uniquely determine the twisted Ishibashi state $| \phi(t, \rho, \vec{x}) \rangle\rangle$, the two states must be identical, and so are the corresponding operators, 
\begin{align}
 \hat{\phi}(t, \rho, \vec{x})=\hat{\phi}^{\mathrm{KLL}}_0(t, \rho,\vec{x}),
\end{align}
by the state-operator correspondence of CFT.

Due to the periodicity of the Green's function in $t$, the KLL state $\hat{\phi}^{\mathrm{KLL}}_0(t, \rho,\vec{x})|0\rangle$ at
$t$ and $t+2\pi$ are identical modulo  
a phase factor $\exp(-2\pi i \Delta_{\phi})$. The twisted Ishibashi state $| \phi(t, \rho, \vec{x}) \rangle\rangle$ has the same 
periodicity since it consists of eigenstates of the global Hamiltonian $H$ with eigenvalues equal to $\Delta_\phi$ plus integers.

When the bulk gravity theory is interacting, 
the leading order KLL operator $\hat{\phi}^{\mathrm{KLL}}_0$ does not satisfy the microscopic causality.
One can incorporate effects due to interactions by modifying the bulk-boundary 
map \eqref{leading} perturbatively 
so that the microscopic causality  is satisfied \cite{Kabat:2011rz, Kabat:2013wga, Heemskerk:2012mn}.
Recently, it was shown in \cite{Kabat:2015swa} that effects in the next leading order in $1/N$ 
can be expressed as a sum of (\ref{leading}) for different primary fields $\phi$, whose coefficients
are determined by the operator product expansion. This procedure can be repeated order by order in
perturbation.  Our result guarantees that these perturbative corrections take the form  (\ref{leading}) 
to all order in perturbation since  we have shown that a bulk local operator is a superposition of twisted Ishibashi operators
provided the compatibility condition (\ref{inf}) is satisfied.  
The superposition coefficients $\psi_\phi$ in  (\ref{superposition}) depend on details of the theory, 
such as the operator product expansion in CFT or the bulk interactions in AdS. 
On the other hand, we may encounter inconsistency between  the microscopic causality and our compatibility requirement (\ref{inf}) 
at a non-perturbative level since we do not expect resolution better than the Planck length in the bulk.

The proposal by  H.~Verlinde \cite{Verlinde:2015qfa} for local operators in AdS$_3$ uses 
Ishibashi states for the full Virasoro symmetry as opposed to the global conformal symmetry and without the
imaginary dilatation $e^{-i \frac{\pi}{2}H}$. In this paper, we have shown that the imaginary dilatation is  
required by the compatibility condition (\ref{inf}), which we think is an essential feature for any bulk local operator
in AdS. Each Ishibashi state for the full Virasoro symmetry can be decomposed into a sum
of those for the global conformal symmetry. Therefore, modulo  the imaginary dilatation, the proposal of  \cite{Verlinde:2015qfa}
can be regarded as a particular choice of the coefficient $\psi_\phi$ in (\ref{superposition}).
With this choice, only global conformal primaries within a single Virasoro representation appear in the superposition.  
However, we expect that  $1/N$ corrections would generate global conformal primaries in other Virasoro representations
also. 

There are several avenues for future investigations.  
The use of boundary states in constructing bulk local operators may be related to 
the computation of the bulk energy density by the Radon transform of the entanglement entropy in 
\cite{Lin:2014hva} and to the integral geometry discussed in \cite{Czech:2015qta}.
We would also like to find out to what extent our proposal depends on the AdS background. 
We note that, for $d=2$ and in the leading order in $1/N$, it has
been pointed out in  \cite{Hamilton:2006fh} that  
the same bulk operator $\hat{\phi}^{\mathrm{KLL}}_0(t, \rho,\vec{x})$  can be used to probe the BTZ black hole 
as well as the pure AdS vacuum. 
It would also be interesting to explore consistency of
the microscopic causality and our compatibility requirement (\ref{inf}) to find out
whether bulk local operators can be defined non-perturbatively in the bulk.

\section*{Acknowledgments}

We thank Tadashi Takayanagi for discussion.
This work is supported in part by U.S.\ DOE grant DE-SC0011632, by 
the Walter Burke Institute for Theoretical Physics, and by the Moore Center for Theoretical Cosmology and 
Physics at Caltech.
The work of H.O.\ is also supported in part by the Simons
Investigator Award, by the WPI Initiative of MEXT of Japan, and by JSPS Grant-in-Aid for Scientific Research C-26400240. 
He also thanks the hospitality of the Aspen Center for Physics, where this paper was completed.
Y.N.\ is a Sherman-Fairchild Research Assistant Professor at Caltech. 
 Though we did not attend the KITP programs, ``Quantum Gravity Foundations: UV to IR" and ``Entanglement in Strongly-Correlated Quantum Matter," we have been benefited by watching their recorded talks.


\end{document}